\documentclass{blois}

\usepackage{wrapfig}

\def\be{\begin{equation}}
\def\ee{\end{equation}}
\def\bea{\begin{eqnarray}}
\def\eea{\end{eqnarray}}

\newcommand{\Photo}{}

\begin{document}
\vspace*{4cm}
\title{Latest results on astroparticle physics with H.E.S.S.}

\author{Lucia Rinchiuso on behalf of the H.E.S.S. Collaboration}

\address{IRFU, CEA, D\'epartement de Physique des Particules, Universit{\'{e}} Paris-Saclay,\\ F-91191 Gif-sur-Yvette, France}

\maketitle
\abstracts{The High Energy Spectroscopic System (H.E.S.S.)  
is pursuing a rich observational program 
in astroparticle physiscs. 
The latest results from H.E.S.S. on the search for dark matter towards the Galactic Center and dwarf galaxies, the test of Lorentz invariance, and the measurement of the cosmic-ray electron spectrum are presented.}

\section{The H.E.S.S. experiment and its astroparticle physics program}
The High Energy Spectroscopic Sysytem (H.E.S.S.) is an array of five Imaging Atmospheric Cherenkov Telescopes (IACT) built in Namibia in 2003. It consists of four telescopes (H.E.S.S.~I) of 12 m diameter  and a 5th larger telescope of 28 m diameter placed in the middle of the array in 2012 (H.E.S.S.~II) in order to lower the energy threshold of the experiment. \\
The H.E.S.S. instrument is able to detect very-high-energy (VHE) $\gamma$-rays in an energy range from a few tens GeV 
to several ten TeV, with an energy resolution 10\% and an angular resolution per $\gamma$ ray better than 0.1$^\circ$. 
The flux sensitivity reaches about 1\% of the Crab flux for 25 h of observations close to zenith. 
The H.E.S.S. collaboration covers a wide range of topics in astroparticle physics. The latest main results focus on dark matter (DM) searches (section~\ref{sec:DM}), searches for Lorentz invariance violation (section~\ref{sec:lorentz}) and the measurement of the VHE cosmic-ray electrons (CRE) spectrum (section~\ref{sec:ele}). 

\section{Dark matter searches}\label{sec:DM}
DM searches can be performed indirectly with H.E.S.S. through the detection of the $\gamma$ rays possibly produced in the final state of the self-annihilation of DM particles. 
The flux of photons from DM annihilation depends on the annihilation channel, the thermally-averaged velocity-weighted annihilation cross section $\langle \sigma v \rangle$, the DM mass and the J-factor. The latter term reflects the DM density distribution in the target. \\
The choice of the best DM targets requires 
large expected DM signal,  
and low standard VHE astrophysical background. 
The H.E.S.S. observational strategy focuses on the region of the Galactic Center (GC, subsection \ref{sub:GC}) and the nearby satellite dwarf spheroidal galaxies of the Milky Way (dSph, subsection \ref{sub:dSph}). 
The DM searches are based on a novel analysis technique that exploits the properties of the DM-induced $\gamma$-ray signals with respect to background. 
The signal is searched in the  {\it ON region} and the background is measured in the {\it OFF region}. If no significant excess is observed in the {\it ON region} a 2 dimensional (2D)-binned (space, energy) log-likelihood ratio test statistic (LLR TS) is performed. The computation of test statistics allows us to derive 95\% confidence level (C.L.) upper limits on $\langle\sigma v\rangle$ as function of the DM mass $m_{\rm DM}$, for a given annihilation channel and DM profile.
\begin{figure*}[ht!]
\centering
\includegraphics[width=0.45\textwidth]{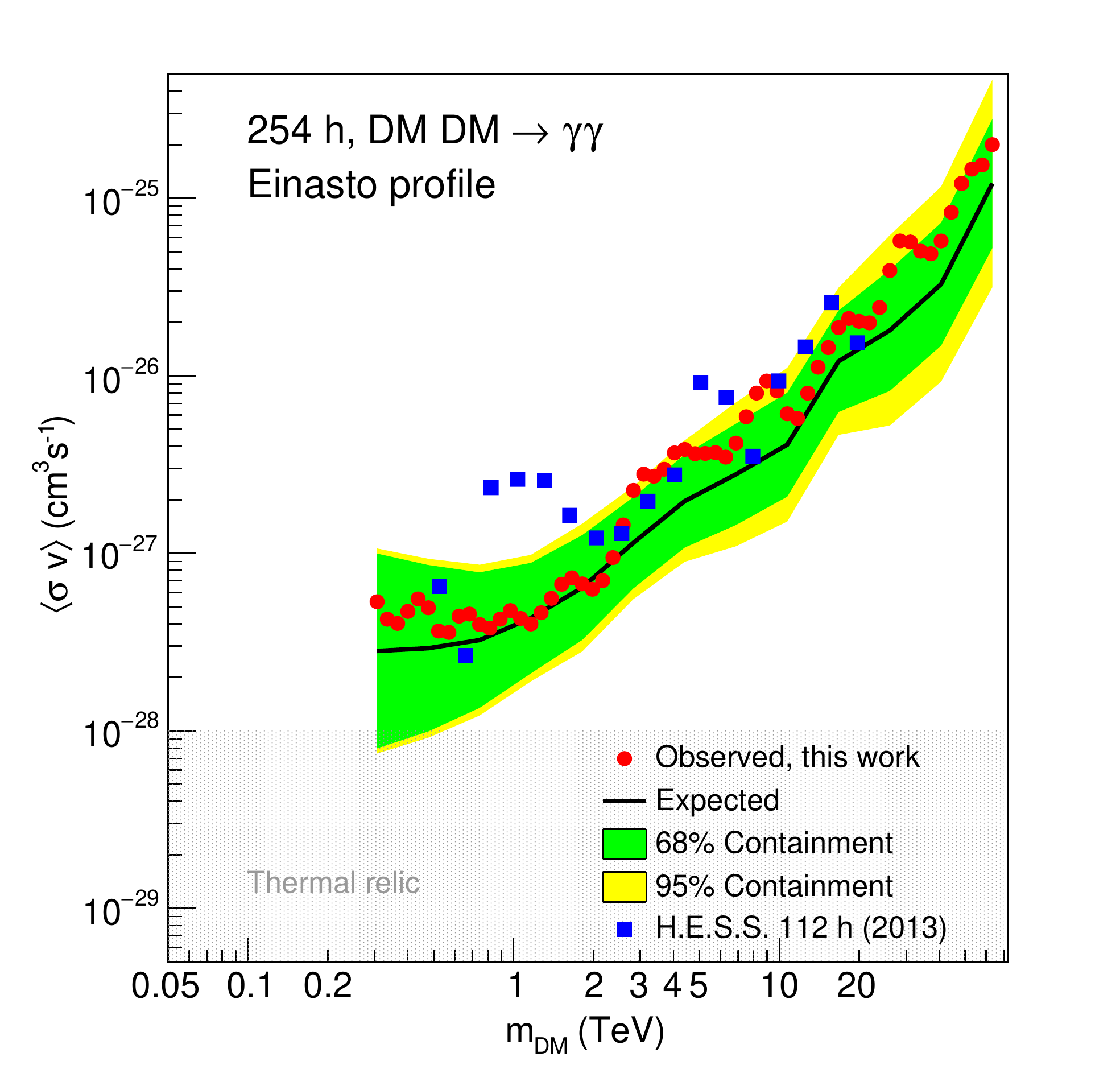}
\includegraphics[width=0.45\textwidth]{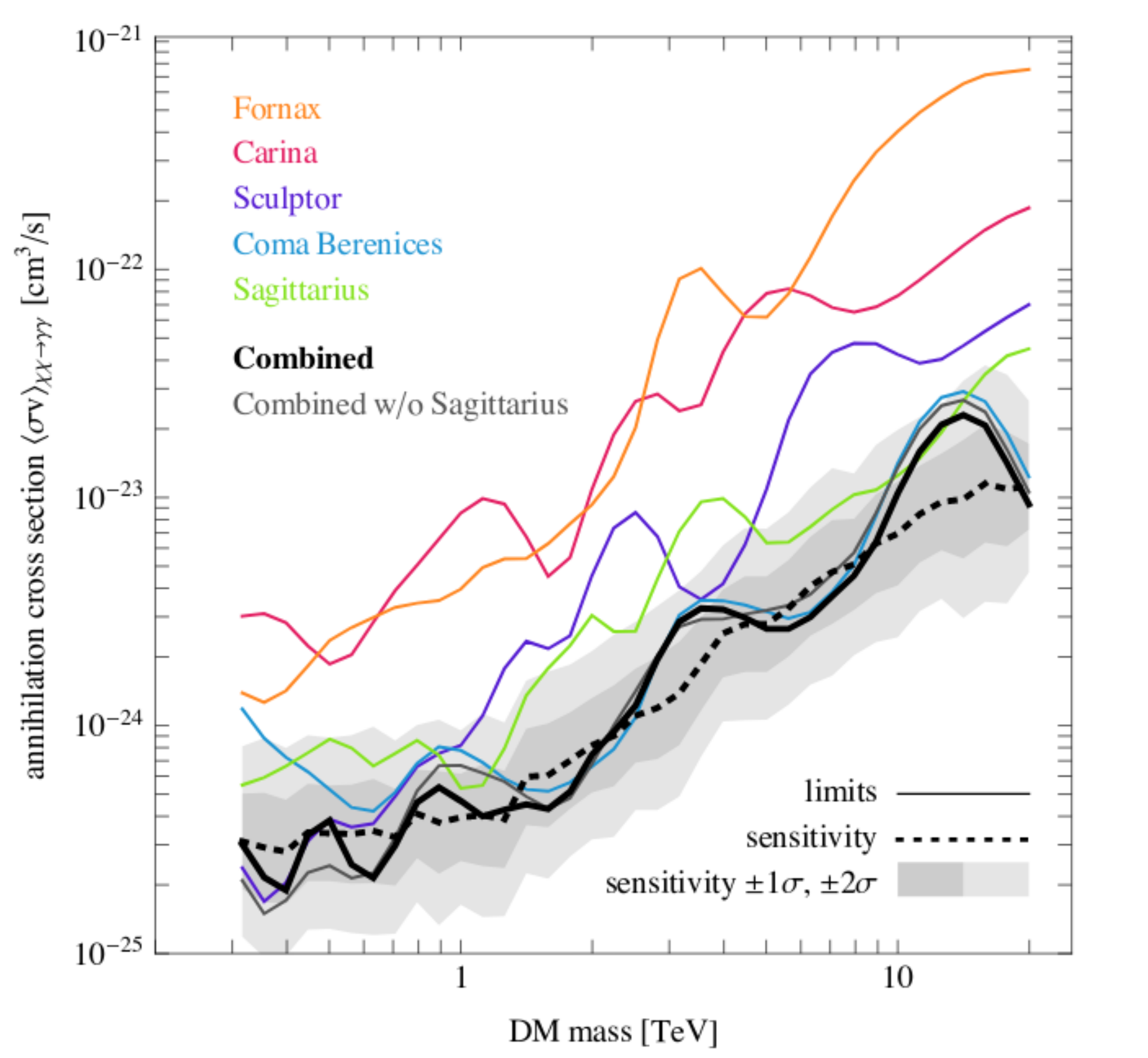}
\caption{95\% C.~L. upper limits from H.E.S.S. I on $\langle \sigma v \rangle$ as a function of the DM mass for the prompt annihilation into two photons. Left: Limits obtained from GC observations~\protect\cite{bib:GC18} assuming an Einasto DM profile. Observed limits (red dots) and mean expected limit (black solid line) together with the 68\% (green band) and 95\% (yellow band) C.~L. containment bands are shown. Left: Individual observed limits for each dSph~\protect\cite{bib:dSph} are show with different colors. The observed (solid black line) and expected (dashed black line) combined limits are shown with their containment bands (gray boxes).}
\label{fig:DM}
\end{figure*}

\subsection{Dark matter at the GC}\label{sub:GC}
The GC region is very promising for DM signals searches due to its proximity and large expected DM content. However, it is very crowded in VHE $\gamma$-rays, including TeV diffuse emissions \cite{bib:Pevatron}, HESS J1745-303\cite{bib:1745-303}, Sagittarius A*\cite{bib:1745-290} and G09+01\cite{bib:G0901}. \\
Because of the pointing strategy and the available photon statistics from H.E.S.S.-I observations, the {\it ON region} is defined as a circle of one degree around the GC and it is split in seven concentric sub regions (RoIs, spatial bins of the likelihood). 
The above-mentioned standard VHE emissions are masked.  
The {\it OFF region} is defined for each RoI and each observation as the symmetric region to the RoI with respect to the pointing position of the observation. This allows us to measure the background in the same observational and instrumental conditions as for the signal. The excluded regions are symmetrically discarded also in the {\it OFF region} in order to obtain {\it ON} and {\it OFF regions} of same solid angle size.  No significant excess is observed between the {\it ON} and {\it OFF region}. \\
The expected DM signal is a monoenergetic $\gamma$-ray line and it is modeled as a Gaussian function of width $\sigma/E$= 10\%. 
95\% C.L. upper limits on $\langle\sigma v\rangle$ as function of $m_{\rm DM}$ are derived with the LLR TS. The observed limits are shown in Fig. \ref{fig:DM} left (red dots) together with the expected limits (black solid line) and their $1\sigma$ and $2\sigma$ containment bands, respectively. The constraints\cite{bib:GC18} reach a value of $4\times10^{-28}$ cm$^3$s$^{-1}$ and improve over previous H.E.S.S. limits 
by a factor about 6 around 1 TeV. H.E.S.S. limits well complement those derived by Fermi-LAT\cite{bib:Fermi}. Together they set the most stringent constraints so far on $\langle\sigma v\rangle$ in the GeV-TeV DM mass range. \\
The H.E.S.S. collaboration is conducting an observational campaign to survey the GC region up to several degrees from Sagittarius A*, which will significantly extend the available photon statistics and allow us to increase the {\it ON region} size to improve the sensitivity for DM searches at at few hundred GeV. 

\subsection{Dark matter toward dSph}\label{sub:dSph}
Nearby dwarf spheroidal galaxy (dSph) satellites of the Milky Way are promising DM targets:  they represent a clean environment for searches of a DM unambiguous signal due to low gas content and absence of recent star formation. H.E.S.S. has carried out a long-term observational campaign on dSphs. A combined analysis has been performed on Carina, 
Coma Berenices, 
Formax, 
Sagittarius 
and Sculptor 
for a total of about 130 hours of observations, with the goal to look for DM 
$\gamma$-lines. \\
The {\it ON region} is a circle centered at the position of the dSph and it is split in 2 (3 for Coma Berenices) RoIs of width $0.1^\circ$. The {\it OFF region} is defined with the ``multiple OFF'' technique: for every RoI and each observation several {\it OFF regions} of same shape and size as the RoI are built at the same distance from the pointing position as for the RoI.\\
No significant excess is observed in any of the dSphs. The 2D-binned likelihood analysis is then applied to compute the 95\% C.L. upper limits on $\sigma v$. Fig.~\ref{fig:DM} right shows the observed limits for the individual dSph dataset and for the combination of the five data sets. Constraints down to $5\times10^{-25}$ cm$^3$s$^{-1}$ are set around 1 TeV thanks to the combination. Observed and expected limits  together with $1\sigma$ and $2\sigma$ containment bands are shown. Specific pure WIMP models have been also tested, using the full spectrum, including the $WW$, $ZZ$, $Z\gamma$ and $\gamma\gamma$ contributions. A significant range of $m_{DM}$ is rejected for the thermal pure WIMP 5tuplet \cite{bib:dSph}. \\
H.E.S.S. has also started to observe some of the recently discovered DES dSphs.

\section{Search for Lorentz invariance violation}\label{sec:lorentz}
Lorentz invariance is a fundamental property of spacetime. However, it could be violated at high energy $E_{\rm QG}$, close to the Planck energy scale $E_{\rm Planck}=1.22\times10^{19}$ GeV, as expected in some quantum gravity models\cite{bib:QG}. This would modify the dispersion relation of photons and introduce an energy dependence of the photon velocity $E^2\simeq p^2c^2\Big[1\pm\big(\frac{E}{E_{\rm QG}}\big)^n\Big]$, where $n$ is the order of the series expansion of the photon dispersion relation ($n=1$ is linear case while  $n=2$ is quadratic) and the sign is negative (positive) for the subluminar (superluminar) case. \\
Among the most promising targets to look for Lorentz invariance violation (LIV) signatures with IATCs are blazars: they are detected up to VHE with high photon statistics and exhibit short-time flux variability. The 2014 flare of Markarian 501 observed by H.E.S.S. up to 20 TeV has been used to look for LIV using two approaches: a temporal study and a spectral study\cite{bib:livMarkarian}. \\
The temporal study is based on a time of flight (ToF) measurement to search for a time lag between two energy bins\cite{bib:LIVtof}  and it had already been applied to other sources observed by H.E.S.S.\cite{bib:livPKS,bib:livPG}. 
In absence of delay, the best estimate of $\tau$ for Markarian 501 is obtained through a likelihood analysis to constrain $E_{\rm QG}$. \\
The LIV-induced photon dispersion relation implies also a modification in the pair production process of VHE $\gamma$-rays on extragalactic background light (EBL) due to starlight and dust re-emission. 
LIV is expected to induce deviations in the spectrum of blazars at VHE energies with respect to the standard EBL pair-production attenuation.
The VHE $\gamma$-ray spectrum of  Markarian 501 flare shows no significant deviation from the standard EBL-attenuated spectrum, which allows us to derive constraints on $E_{\rm QG}$. The EBL model from Ref.\cite{bib:EBLmodel} is assumed.
The 95\% C.L. lower limits are summarized in Tab. \ref{tab:LIV}. The constraints on the linear case from the spectral study probe the Planck scale. $E_{\rm QG} = 2.1E_{\rm Planck}$ is excluded at $5.8\sigma$.
The constraints obtained with Markarian 501 flare are among the most stringent obtained so far. 
\begin{table}[h]
\centering
\begin{tabular}{c | c | c}
\hline\hline
n & ToF subluminal & Spectrum \\
\hline\hline
1 & $2.71\times10^{11}$ GeV &  $2.56\times10^{19}$ GeV \\
2 & $7.34\times10^{10}$ GeV  & $7.81\times10^{11}$ GeV \\
\hline
\hline
\end{tabular}
\caption{95\% C.L. lower limits on $E_{QG}$ for the linear (n=1) and quadratic case (n=2), extracted through the temporal and spectral analyses, respectively, for the subluminal case.}
\label{tab:LIV}
\end{table}

\section{Cosmic-ray electron spectrum measurement}\label{sec:ele}
The VHE cosmic-ray electron (CREs) spectrum is a crucial probe to study the properties of the electron sources and electron propagation in the interstellar medium.  
Since VHE electrons suffer from severe Inverse Compton and synchrotron energy losses, CREs can only come from  a handful of nearby sources. TeV electrons are expected to be produced in sources within about 1 kpc from the Sun. \\
In the H.E.S.S. data set CREs are selected among the residual background measured in observations at several degrees from the Galactic plane to avoid VHE $\gamma$-ray sources and diffuse emission. 
The new H.E.S.S. spectrum\cite{bib:CReHESS} improves over the previous measurement from increased CRE statistics 
and improved event selection (better hadron rejection).
The H.E.S.S. spectrum is best fitted with a broken power law with photon indexes 3.04 at low energies and 3.78 at higher above the energy break. The energy break at $\sim1$ TeV is compatible with previous measurements. 
A careful treatment of the sources of systematic uncertainties is carried out including the proton contamination, year-by-year variation of the CRE spectrum, dependency on the zenith angle and variation of atmospheric conditions. 
In particular, the CRE spectrum below 5 TeV is in very good agreement  with the recent measurements by VERITAS\cite{bib:CReVeritas}. 
No significant spectral feature is observed, which can be used to put constraints on propagation models ({\it i.e.} estimate of diffusion coefficient) and specific sources such as nearby pulsars or DM clumps.

\section{Conclusions}
H.E.S.S. carries out a wide program in astroparticle  physics that provides strong constraints on DM searches towards the GC and dSph galaxies and on LIV searches, and the first CRE spectrum measurement in the ten TeV energy range. The observational program at the GC is pursued with a survey in the inner galaxy region with observations 
 at larger Galactic latitudes from the GC. This will significantly enlarge the present data set and the searched region to improve the sensitivity of DM signal searches. Joint projects between the  H.E.S.S., MAGIC and VERITAS collaborations have started on LIV studies and a DM search towards dSph galaxies in order to combine the data sets on different targets. The aim is to exploit the complementarity of the experiments to improve the current knowledge about some of the major open topics in astroparticle physics.

\end{document}